\documentstyle[fleqn]{article}
\def\beq{\begin{equation}}
\def\eeq{\end{equation}}
\def\Eq{Eq.~(\ref}
\def\0{\otimes}
\def\6{\langle}
\def\9{\rangle}
\def\Tr{\mbox{Tr}\,}
\def\bc{\medskip\begin{center}}
\def\ec{\end{center}\medskip}
\def\cC{$\cal C$}

\begin{document}

\renewcommand{\thefootnote}{\fnsymbol{footnote}}

\vspace*{\fill}
\begin{center}
{\Large {\bf Classical interventions in quantum systems.\medskip 

I. The measuring process}}\\[15mm]

Asher Peres\footnote{E-mail: peres@photon.technion.ac.il} \\[8mm]
{\sl Department of Physics, Technion---Israel Institute of Technology,
32\,000 Haifa, Israel}\\[15mm]

{\bf Abstract}

\end{center}

The measuring process is an external intervention in the dynamics of a
quantum system. It involves a unitary interaction of that system with a
measuring apparatus, a further interaction of both with an unknown
environment causing decoherence, and then the deletion of a subsystem.
This description of the measuring process is a substantial
generalization of current models in quantum measurement theory. In
particular, no ancilla is needed. The final result is represented by a
completely positive map of the quantum state $\rho$ (possibly with a
change of the dimensions of $\rho$). A continuous limit of the above
process leads to Lindblad's equation for the quantum dynamical
semigroup.\vfill

\noindent PACS numbers: 03.65.Bz, \ 03.67.* \vfill

\noindent Physical Review A {\bf61}, 022116 (2000) \vfill

\newpage \begin{center}{\bf I. INTRODUCTION}\ec

The measuring process~\cite{vN,WZ} is the interface of the classical and
quantum worlds. The classical world has a description which may be
probabilistic, but in a way that is compatible with Boolean logic. In
the quantum world, probabilities result from complex amplitudes that
interfere in a non-classical way. In this article, the notion of
measurement is extended to a more general one: an {\it intervention\/}.
An intervention has two consequences. One is the acquisition of
information by means of an apparatus that produces a record. This step
is called a {\it measurement\/}. Its outcome, which is in general
unpredictable, is the {\it output\/} of the intervention. The other
consequence is a change of the environment in which the quantum system
will evolve after completion of the intervention. For example the
intervening apparatus may generate a new Hamiltonian that depends on the
recorded result. In particular, classical signals may be emitted for
controlling the execution of further interventions. In the second part
of this article~\cite{II}, these signals will be limited to the velocity
of light, so as to obtain a relativistic version of quantum measurement
theory.

Interventions are mathematically represented by completely positive
maps. Their properties are discussed in Sect.~II, where a detailed
dynamical description is given of the measuring process: it involves
unitary interactions with a measuring apparatus and with an unknown
environment that causes decoherence, and then the optional deletion of a
subsystem. The Hilbert space for the resulting quantum system may have a
different number of dimensions than the initial one. Thus, a quantum
system whose description starts in a given Hilbert space may evolve into
a set of Hilbert spaces with different dimensions. If one insists on
keeping the same Hilbert space for the description of the entire
experiment, with all its possible outcomes, this can still be achieved
by defining it as a Fock space.

The term ``detector'' will frequently appear in this article and in the
following one. It means an elementary detecting element, such as a
bubble in a bubble chamber, or a small segment of wire in a wire
chamber. Note that in such a detector, the time required for the
irreversible act of amplification (the formation of a microscopic
bubble, or the initial stage of the electric discharge) is extremely
brief, typically of the order of an atomic radius divided by the
velocity of light. Once irreversibility has set in, the rest of the
amplification process is essentially classical. It is noteworthy that
the time and space needed for initiating the irreversible processes are
incomparably smaller than the macroscopic resolution of the detecting
equipment.

An intervention is described by a set of parameters that include the
time at which the intervention occurs. Interventions of finite duration
can also be considered~\cite{PW85} and will be briefly discussed. For
a relativistic treatment (in the following article), we shall need the
location of the intervention in spacetime, referred to an arbitrary
coordinate system. In any case, we have to specify the speed and
orientation of the apparatus in the coordinate system that we are using
and various other {\it input\/} parameters that control the apparatus,
such as the strength of a magnetic field, or that of an RF pulse used in
the experiment, and so on. The input parameters are determined by
classical information received from past interventions, or they may be
chosen arbitrarily by the observer who prepares that intervention, or by
a local random device acting in lieu of the observer.

A crucial physical assumption is that there exists an objective time
ordering of the various interventions in an experiment. There are no
closed causal loops. This time ordering defines the notions earlier and
later. The input parameters of an intervention are deterministic (or
possibly stochastic) functions of the parameters of earlier
interventions, but not of the stochastic outcomes resulting from later
interventions. In such a presentation, there is no ``delayed choice
paradox''~\cite{delayed} (there can be a delayed choice, of course, but
no paradox is associated with it).

The word ``measurement'' is a bit misleading, because it suggests that
there exists in the real world some unknown property that we are
measuring~\cite{qt}. This term was banned by Bell~\cite{Bell}, though
for a different reason: Bell pointed out that the notion of measurement,
or observation, was logically inconsistent in a world whose description
is purely quantum mechanical. However, the approach followed in the
present article does not comply with Bell's desiderata. It explicitly
associates classical inputs and outputs with each intervention
\cite{BJ,percival}.

The probabilities of the various outcomes of an intervention can be
predicted by using a suitable theory, such as quantum theory. Besides
these outcomes, there may also be other output parameters: there may
be modifications of the physical environment depending on which outcome
arose, and the intervening apparatus may emit classical signals with
instructions for setting up later interventions. As a concrete example,
consider the quantum teleportation scenario~\cite{telep}. The first
intervention is performed by Alice: she has two spin-$1\over2$ particles
and she performs on them a test with four possible outcomes. When Alice
gets the answer, she emits a corresponding signal, which becomes an
input for Bob's intervention: the latter is one of four unitary
transformations that can be performed on Bob's particle. 

Quantum mechanics is fundamentally statistical~\cite{Bal}. In the
laboratory, any experiment has to be repeated many times in order to
infer a law; in a theoretical discussion, we may imagine an infinite
number of replicas of our gedanken\-experiment, so as to have a genuine
statistical ensemble. The various experiments that we consider all start
in the same way, with the same initial state $\rho_0$, and the first
intervention is the same. However, later stages of the experiment may
involve different types of interventions, possibly with different
spacetime locations, depending on the outcomes of the preceding events.
Yet, assuming that each intervention has only a finite number of
outcomes, there is for the entire experiment only a finite number of
possible records. (Here, the word ``record'' means the complete list of
outcomes that occurred during the experiment. I do not want to use the
word ``history'' which has acquired a different meaning in the writings
of some quantum theorists.)

Each one of these records has a definite probability in the statistical
ensemble. In the laboratory, experimenters can observe its relative
frequency among all the records that were obtained; when the number of
records tends to infinity, this relative frequency is expected to tend
to the true probability. The role of theoretical physics is to predict
the probability of each record, given the inputs of the various
interventions (both the inputs that are actually controlled by the local
experimenter and those determined by the outputs of earlier
interventions). Note that each record is objective: everyone agrees on
what happened (e.g., which detectors clicked). Therefore, everyone
agrees on what the various relative frequencies are, and the theoretical
probabilities are also the same for everyone.

The ``detector clicks'' are the only real thing we have to consider.
Their observed relative frequencies are objective data. On the other
hand, wave functions and operators are nothing more than abstract
symbols. They are convenient mathematical concepts, useful for computing
quantum probabilities, but they have no real existence in
Nature~\cite{Stapp}. Note also that while interventions are localized in
spacetime, quantum systems are pervasive. In each experiment,
irrespective of its history, there is only one quantum system. The
latter typically consists of several particles or other subsystems, some
of which may be created or annihilated at the various interventions.

The next section describes the quantum dynamics of the measuring process
which is an essential part of each intervention. The role of
decoherence, due to an inavoidable interaction with an unknown
environment, is discussed in Sect.~III. The final result, \Eq{ArhoA}),
will be extensively used in the following article. The right hand side
of that equation contains operators $A_{\mu m}$ which are typically
represented by rectangular matrices. Some of their mathematical
properties (in particular factorability) are discussed in Sect.~IV.

Decoherence, whose role is essential in the measuring process, is a
stochastic phenomenon similar to Brownian motion. However, when seen on
a coarse time scale, it is possible to consider it as a continuous
process. This continuous approximation leads to the Lindblad
equation~\cite{lindblad}, which is derived in a simple and novel way in
Sect.~V.

\bc {\bf II. THE MEASURING PROCESS}\ec

The measuring process involves several participants: the physical system
under study, a measuring apparatus whose states belong to
macroscopically distinguishable subspaces, and the ``environment'' which
includes unspecified degrees of freedom of the apparatus and the rest of
the world. These unknown degrees of freedom interact with the relevant
ones, but they are not under the control of the experimenter and cannot
be explicitly described. Our partial ignorance is not a sign of
weakness. It is fundamental. If everything were known, acquisition of
information would be a meaningless concept.

In order to keep the discussion as general as possible, I do not
introduce here any ``ancilla,'' contrary to current fashion. This
omission is not an oversight. It is intentional and deserves a brief
explanation. In the early years of quantum mechanics, von Neumann wrote
a rigorous mathematical treatise~\cite{vN} which had a lasting
influence. According to von Neumann, the various outcomes of a
measurement correspond to a complete set of orthogonal projection
operators in the Hilbert space of the quantum system under study. It was
later realized that von Neumann's approach was too restricted, because
the measuring process may have more distinct outcomes than the number of
dimensions of that Hilbert space. The appropriate formalism is that of a
{\it positive operator valued measure\/} (POVM)~\cite{JP,DL}. Namely,
the various outcomes of the measurement correspond to positive operators
$E_\mu$, which sum up to the unit operator but need not commute. 

This raised a new problem: the actual implementation of a given POVM. In
the final section of his book, von Neumann had formally shown how to
construct a Hamiltonian that generated a dynamical evolution of the type
required to obtain a projection valued measure (PVM). This was a
mathematical proof of existence, namely, quantum dynamics was compatible
with the structure of a PVM. Is it compatible with a more general POVM?
This question was answered by Helstrom~\cite{Hel} who converted the
problem of implementation of a POVM into that of an ordinary von Neumann
measurement, by introducing an auxiliary quantum system that he called
{\it ancilla\/} (the Latin word for housemaid). By virtue of Neumark's
theorem~\cite{Neu}, any POVM can be obtained from a PVM applied to a
composite system that consists of the original system and an ancilla
having a sufficient number of dimensions. This provides a formal proof
of existence, but in real life this is usually not how measurements are
actually performed. Anyway, even if an ancilla is used according to
Helstrom's protocol, we may as well consider it as part of the measuring
apparatus. Therefore the following description of the measuring process
will not involve any ancilla, and yet it will explicitly show how any
POVM can be implemented by a unitary interaction of the quantum system
with a suitable apparatus.

To simplify the notations, it will be assumed that finite dimensional
Hilbert spaces are sufficient for describing the quantum system under
study, the apparatus, and even the environment. Moreover, the initial
states $\rho_i$ of the system and the apparatus are assumed pure.
Initially mixed states would be a more realistic assumption, but since
they can always be written as convex combinations of pure $\rho_i$,
their use would not bring any essential change in the discussion below.

Let a set of basis vectors for the system under study be denoted as
$\{|s\9\}$. The initial state of that system is a linear combination,
$|\psi_0\9=\sum c_s|s\9$, with complex coefficients $c_s$. Let $|A\9$ be
the initial state of the apparatus. In the first step of the measuring
process, which may be called a ``premeasurement''~\cite{undo}, the
apparatus interacts unitarily with the quantum system and becomes
entangled with it, so that they effectively become a single composite
system \cC:

\beq \sum_s c_s\,|s\9\0|A\9\to
  \sum_{s,\lambda} c_s\,U_{s\lambda}\,|\lambda\9,\label{premea}\eeq
where $\{|\lambda\9\}$ is a complete basis for the states of \cC. It is
the choice of $U_{s\lambda}$ that determines which property of the
system under study is correlated to the apparatus, and therefore is
measured. Unitarity implies that

\beq \sum_\lambda U_{s\lambda}\, U_{t\lambda}^*=\delta_{st},
 \label{unitU}\eeq
or $UU^\dagger={\bf1}$, where {\bf1} is the unit matrix in the Hilbert
space of the original quantum system. The $U$ matrix is not square: it
has fewer rows than columns, because we have considered only a single
initial state of the apparatus, namely $|A\9$. If we had introduced a
complete set of states for the apparatus, labelled $|B\9$, $|C\9$, etc.,
then $U$ would have been a unitary matrix satisfying

\beq \sum_\lambda U_{sA,\lambda}\, U_{tB,\lambda}^*=
 \delta_{st}\,\delta_{AB}.\eeq
Our freedom of choosing the required unitary matrix $U_{sA,\lambda}$ is
equivalent to the freedom of choice of an interaction Hamiltonian in the
von Neumann formalism.

The apparatus itself is an utterly complicated system, and some radical
assumptions are needed in order to proceed with explicit calculations.
The assumptions below are not as drastic as those commonly used in
quantum measurement theory, yet they ought to be clearly spelled out.
Let us assume that the composite system \cC\ can be fully described by
the theory. Its complete description involves both ``macroscopic''
variables and ``microscopic'' ones. The difference between them is that
the microscopic degrees of freedom can be considered as adequately
isolated from the environment for the duration of the experiment, so
that their evolution is in principle perfectly controlled, while the
macroscopic ones cannot be isolated from the unknown environment and
the dynamical evolution cannot be completely predicted. Statistical
hypotheses are required in order to make plausible predictions, as
explained below. Any other degrees of freedom of the apparatus, for
which no explicit description is provided, are considered as part of the
environment.

An essential property of the composite system \cC, which is necessary to
produce a meaningful measurement, is that its states form a finite
number of orthogonal subspaces which are distinguishable by the
observer. These subspaces are similar to, but more general than, Zurek's
``pointer basis''~\cite{pointer} which is a preferred basis for the
apparatus. Here, we consider ortho\-gonal {\it subspaces\/} of the
composite system \cC, which may have different numbers of dimensions.
For example, a particle detector may have just two such subspaces:
``ready to fire'' and ``discharged'' (obviously, the latter has many
more states than the former). Each macroscopically distinguishable
subspace corresponds to one of the outcomes of the intervention and
defines a POVM element $E_\mu$, given explicitly by \Eq{povm}) below.
The labels $\mu$ are completely arbitrary, for example they may be the
labels printed on the various detectors. The initial state of \cC,
namely $|\psi_0\9\0|A\9$, lies in the subspace that corresponds to the
null outcome (no detection).

Obviously, the number of different outcomes $\mu$ is far smaller than
the dimensionality of the composite system \cC. Let us introduce a
complete basis, $\{|\mu,\xi\9\}$, where $\mu$ labels a macroscopic
subspace, as explained above, and $\xi$ labels microscopic states in
that subspace. We thus have

\beq \6\mu,\xi|\nu,\eta\9=\delta_{\mu\nu}\,\delta_{\xi\eta}.\eeq
Note that the various subspaces labelled $\mu$ may have different
dimensions, that is, the range of indices $\xi$ may depend on the
corresponding $\mu$. We shall henceforth write $U_{s\mu\xi}$ instead of
$U_{s\lambda}$.

After the premeasurement, given by \Eq{premea}), the state of \cC\ is
given by

\beq |\psi_1\9=\sum_{s,\mu,\xi} c_s\,U_{s\mu\xi}\,|\mu,\xi\9.
 \label{psi1} \eeq
The probability of obtaining outcome $\mu$ is the contribution of
subspace $\mu$ to the density matrix $|\psi_1\9\6\psi_1|$. Explicitly,
it is

\beq \sum_{s,t,\xi} c_sc_t^*\,U_{s\mu\xi}\,U^*_{t\mu\xi}=
  \sum_{s,t}\rho_{st}\,(E_\mu)_{ts}, \label{prob}\eeq
where

\beq (E_\mu)_{ts}=\sum_\xi U_{s\mu\xi}\,U^*_{t\mu\xi} \label{povm}\eeq
is a POVM element, defined in the Hilbert space of the system under
study whose initial state was $\rho_{st}$. Note that the matrices
$E_\mu$ satisfy

\beq \sum_\mu(E_\mu)_{ts}=\sum_{\mu,\xi} U_{s\mu\xi}\,U^*_{t\mu\xi}=
 \delta_{ts}, \eeq
by virtue of the unitarity property in \Eq{unitU}).

\bc {\bf III. DECOHERENCE}\ec

Up to now, the quantum evolution is well defined and it is in principle
reversible. It would remain so if the macroscopic degrees of freedom of
the apparatus could be perfectly isolated from their environment, and in
particular from the ``irrelevant'' degrees of freedom of the apparatus
itself. This demand is of course self-contradictory, since we have to
read the result of the measurement if we wish to make any use of it.

Let $\{|e_\alpha\9\}$ denote a complete basis for the states of the
environment, and let $|e_\omega\9$ be the state of the environment at
the moment of the premeasurement. That state is of course unknown, but I
temporarily assume that it is pure, and moreover that it is one of the
basis states, in order to simplify the notations. This pure initial
state will later be replaced by a density matrix,

\beq \rho_e=\sum_\omega p_\omega\,|e_\omega\9\6e_\omega|,
 \label{rhoenv} \eeq
with unknown random non-negative coefficients $p_\omega$. There is no
loss of generality in assuming that $\rho_e$ is diagonal in the basis
used for the states of the environment. This merely says that this basis
was chosen in the appropriate way.

Recall that states $|\mu,\xi\9$ with different $\mu$ are macroscopically
different, so that they interact with different environments. On the
other hand, the labels $\xi$ refer to micro\-scopic degrees of freedom
that are well protected from parasitic disturbances. This sharp
dichotomous distinction between the two types of degrees of freedom is
the only approximation that was made until now.

The unitary interaction of \cC\ with the environment thus generates an
evolution which is not under the control of the experimenter:

\beq |\mu,\xi\9\0|e_\omega\9\to
 |\mu,\xi\9\0\sum_\alpha b_{\mu\omega\alpha}|e_\alpha\9.
 \label{deco}\eeq
The coefficients $b_{\mu\omega\alpha}$ are unknown except for
normalization,

\beq \sum_\alpha |b_{\mu\omega\alpha}|^2=1. \label{bnorm}\eeq
They have no subscript $\xi$ because the microscopic degrees of freedom
do not interact with the environment, and they cannot mix different
values of $\mu$ because the latter refer to macroscopic outcomes that
are stable on the time scale of the experiment. One could also consider
a more general evolution, where the right hand side of \Eq{deco}) would
involve different subspaces $\mu$. This would mean that the measuring
apparatus is actually disturbed by the environment. Such a process is
called {\it noise\/} and is essentially different from the phenomenon of
{\it decoherence\/}, whose occurence is explained below. Here it is
assumed that no noise affects the measuring process. It's only the
environment, whose microscopic degrees of freedom are not robust, that
is disturbed by the apparatus (this is the mechanism causing
decoherence). Exactly how it is disturbed cannot be known, however we do
know that macroscopically different states of \cC\ lead to different
disturbances of the environment, hence the appearance of an index $\mu$
in the coefficients $b_{\mu\omega\alpha}$.

The final state with all the participating subsystems thus is

\beq |\psi_2\9=\sum_{s,\mu,\xi,\alpha}c_s\,U_{m\mu\xi}\,
  b_{\mu\omega\alpha}\,|\mu,\xi\9\0|e_\alpha\9. \eeq
The final density matrix (still a pure state, for the sake of simple
notations) is $\rho=|\psi_2\9\6\psi_2|$. Explicitly, we see from
Eq.~(\theequation) that the expression for $\rho$ contains, among other
things, operators $|e_\alpha\9\6e_\beta|$ which refer to states of the
environment. They are unknown and are considered as unknowable. The only
operator acting on these states that we know to write is represented by
the unit matrix $\delta_{\alpha\beta}$. Its meaning, in the laboratory,
is that of complete ignorance. Therefore we can effectively replace the
complete density matrix $\rho$ by the reduced matrix obtained from it by
ignoring the inaccessible degrees of freedom of the environment. That
is, we replace the operators $|e_\alpha\9\6e_\beta|$ which appear in
$\rho$ by $\6e_\beta|e_\alpha\9=\delta_{\alpha\beta}$, and we perform a
partial trace on the indices that refer to the environment.

The reduced density matrix thus contains expressions of the type
$\sum_\alpha b_{\mu\omega\alpha}\,b^*_{\nu\omega\alpha}$. It will now be
shown that, after a reasonably short time has elapsed, we have

\beq \sum_\alpha b_{\mu\omega\alpha}\,b^*_{\nu\omega\alpha}\simeq
  \delta_{\mu\nu}. \label{decohe} \eeq
The case $\mu=\nu$ is the normalization condition (\ref{bnorm}) due to
unitarity. When $\mu\neq\nu$, the rationale for arguing that the left
hand side of (\theequation) is very close to zero (and has a random
phase) is that the environment has a huge number of states, $N$ say,
whose dynamics is chaotic. Therefore the scalar product of any two
states such as $\sum b_{\mu\omega\alpha}|e_\alpha\9$ and $\sum
b_{\nu\omega\beta}|e_\beta\9$ that may result from \Eq{deco}), at any
random time, is of the order of $N^{-1/2}$, because the components of a
random state, in a randomly chosen basis, are of the order of
$N^{-1/2}$. The time that has to elapse to make Eq.~(\theequation) a
good approximation is called the {\it decoherence time\/}, and it
depends on how well the macroscopic degrees of freedom of the measuring
apparatus are isolated from the environment. There may of course be
large fluctuations on the left hand side of Eq.~(\theequation), akin to
Poincar\'e recurrences~\cite{schul}, but this expression is {\it almost
always\/} very close to zero if $\mu\neq\nu$.

The approximation becomes even better if instead of an ideal pure
initial state $|e_\omega\9$ for the environment, we take the more
realistic density matrix given by \Eq{rhoenv}). Instead of
Eq.~(\theequation) we then have

\beq \sum_\omega p_\omega \sum_\alpha b_{\mu\omega\alpha}\,
  b^*_{\nu\omega\alpha}\simeq \delta_{\mu\nu}, \eeq
where the off-diagonal terms on the right hand side now are of order
$N^{-1}$, rather than $N^{-1/2}$ as before. It is plausible that the
above argument can be made mathematically rigorous in the thermodynamic
limit $N\to\infty$.

It follows that states of the environment that are correlated to
subspaces of \cC\ with different labels $\mu$ can safely be treated in
our calculations as if they were exactly orthogonal. The resulting
theoretical predictions will almost always be correct, and if any rare
small deviation from them is ever observed, it will be considered as a
statistical quirk, or an experimental error. The reduced density matrix
thus is block-diagonal and all our statistical predictions are identical
to those obtained for an ordinary mixture of (unnormalized) pure states

\beq |\psi_\mu\9=\sum_{s,\xi}c_s\,U_{s\mu\xi}\,|\mu,\xi\9,
 \label{psimu} \eeq
where the statistical weight of each state is the square of its norm.
This mixture replaces the pure state $|\psi_1\9$ in \Eq{psi1}). This is
the meaning of the term decoherence. From this moment on, the
macroscopic degrees of freedom of \cC\ have entered into the classical
domain~\cite{Bohr1939,hay}. We can safely observe them and ``lay on them
our grubby hands''~\cite{caves}. In particular, they can be used to
trigger amplification mechanisms (the so-called detector clicks) for the
convenience of the experimenter. 

Note that all these properties still hold if the measurement outcome
happens to be the one labelled $\mu=0$ (that is, if there is no detector
click). It does not matter whether this is due to an imperfection of the
detector or to a probability less than~1 that a perfect detector would
be excited. The state of the quantum system does not remain unchanged.
It has to change to respect unitarity. The mere presence of a detector
that could have been excited implies that there has been an interaction
between that detector and the quantum system. Even if the detector has a
finite probability of remaining in its initial state, the quantum system
correlated to the latter acquires a different state~\cite{Dicke}. The
absence of a click, when there could have been one, is also an event and
is part of the historical record.

The final (optional) step of the intervention is to discard part of the
composite system \cC. In the case of a von Neumann measurement, the
subsystem that is discarded and thereafter ignored is the measuring
apparatus itself. In general, it is a different subsystem: the discarded
part may depend on the outcome $\mu$ and in particular its dimensions
may depend on $\mu$. The remaining quantum system then also has
different dimensions. We therefore introduce in the subspace $\mu$ two
sets of basis vectors $|\mu,\sigma\9$ and $|\mu,m\9$ for the new system
and the part that is discarded, respectively. They replace the original
basis $|\mu,\xi\9$ and it is convenient to choose the latter in such a
way that for each $\xi$ we can write

\beq |\mu,\xi\9=|\mu,\sigma\9\0|\mu,m\9 \eeq
as a direct product, rather than a linear superposition of such
products.

We are now ready to discard the subsystem whose basis vectors are
denoted as $|\mu,m\9$. In the unnormalized density matrix $\rho_\mu=
|\psi_\mu\9\6\psi_\mu|$ (whose trace is the probability of observing
outcome $\mu$), we ignore the deleted subsystem. Namely, we replace the
operator $|\mu,m\9\6\mu,n|$ that appears in $\rho_\mu$ by a unit matrix
$\delta_{mn}$ and we perform a partial trace on the indices $m$ and $n$,
as we have done when we discarded the states of the environment. We thus
obtain a reduced density matrix

\beq \rho'_\mu=\sum_{s,t}c_sc_t^*\sum_{m,\sigma,\tau}U_{s\mu\sigma m}\,
 U^*_{t\mu\tau m}\,|\mu,\sigma\9\6\mu,\tau|. \eeq
Its elements $\6\mu,\sigma|\rho'_\mu|\mu,\tau\9$ can be written as

\beq (\rho'_\mu)_{\sigma\tau}=\sum_{m}\sum_{s,t}
 (A_{\mu m})_{\sigma s}\;\rho_{st}\;(A_{\mu m}^*)_{\tau t},
 \label{rhomu} \eeq
where $\rho_{st}\equiv c_sc_t^*$, and the notation

\beq (A_{\mu m})_{\sigma s}\equiv U_{s\mu\sigma m}, \label{AU}\eeq
was introduced for later convenience. Recall that the indices $s$ and
$\sigma$ refer to the original system under study and to the final one,
respectively. Omitting these indices, \Eq{rhomu}) takes the familiar
form

\beq \rho'_\mu=\sum_{m} A_{\mu m}\,\rho\,A_{\mu m}^\dagger,
 \label{ArhoA} \eeq
which is the most general completely positive linear map~\cite{choi}.
This is sometimes written as $\rho'_\mu=S\rho$, where $S$ is a linear
{\it super\-operator\/} which acts on density matrices (while an
ordinary operator acts on quantum states). Note however that these
super\-operators have a very special structure, given by
Eq.~(\theequation).

Clearly, the ``quantum jump'' $\rho\to\rho'_\mu$ is not a dynamical
process that occurs in the quantum system by itself. It results from the
introduction of an apparatus, followed by its deletion or that of
another subsystem. In the quantum folklore, an important role is played
by the ``irreversible act of amplification.'' The latter is irrelevant
to the present issue. The amplification is solely needed to help the
experimenter. A jump in the quantum state occurs even when there is no
detector click or other macroscopic amplification, because we impose
abrupt changes in our way of delimiting the object we consider as the
quantum system under study. The precise location of the intervention,
which is important in a relativistic discussion~\cite{II}, is the point
from which classical information is sent that may affect the input of
other interventions. More precisely, it is the earliest spacetime point
from which classical information could have been sent. This is also true
for interventions that gave no detection event. Such a passive
intervention is located where the detection event would have occurred,
if there had been one.

Is it possible to maintain a strict quantum formalism and treat the
intervening apparatus as a quantum mechanical system, without ever
converting it to a classical description? We could then imagine not only
sets of apparatuses spread throughout spacetime, but also truly
delocalized apparatuses~\cite{AA80}, akin to Schr\"odinger cats
\cite{monroe,placek}, so that interventions would not be localized in
spacetime as required in the present formalism. However, such a process
would only be the creation of a correlation between two nonlocal quantum
systems. This would not be a true measurement but rather a
``premeasurement''~\cite{undo}. A valid measuring apparatus must admit a
classical description equivalent to its quantum description~\cite{hay}
and in particular it must have a positive Wigner function. Therefore a
delocalized apparatus is a contradiction in terms. If a nonlocal system
is used for the measurement, it must be described by quantum mechanics
(no classical description is possible), and then it has to be measured
by a valid apparatus that behaves quasi-classically and in particular is
localized. It can indeed be localized as well as we wish, if it is
massive enough.

Likewise, quantum measurements of finite duration, as discussed by Peres
and Wootters~\cite{PW85}, actually are only premeasurements. To obtain
consistent results, these authors had to explicitly introduce a second
apparatus that suddenly measures the first one. Their first apparatus
has no classical description. In the language that I am using now, only
the second apparatus performs a valid measurement.

In a purely quantum description of the apparatus, which is the one
appropriate at the premeasurement stage, the new state is an incoherent
mixture of various $\rho_\mu$ correlated to distinct outcomes $\mu$ of
the apparatus. However, the description of the apparatus must ultimately
be converted into a classical one~\cite{Bohr1939,hay} if we want it to
yield a definite record. On the other hand, it is also possible to
discard the apparatus {\it without\/} recording its result. We then have
to describe the state of the quantum system by a mixture of mixtures, as
in \Eq{noselect}) below. The term ``compound''~\cite{marlow} has been
proposed for that kind of mixture which is solely due to our ignorance
of the actual outcome and has no objective nature. Once we have a
definite outcome $\mu$, the new state is $\rho_\mu$, given by
\Eq{ArhoA}).\clearpage

\bc {\bf IV. KRAUS MATRICES}\ec

A special case of Eq.~(\theequation) for square matrices $A_{\mu m}$
was obtained by Kraus~\cite{Kraus} who sought to find the most general
completely positive map for the density matrices of a given quantum
system (no change of dimensions was allowed). Kraus's result obviously
is a generalization of von Neumann's prescription for the state
resulting from the $\mu$-th outcome of a measurement, namely
$\rho_\mu=P_\mu\rho P_\mu$, where $P_\mu$ is the projection operator
associated with outcome $\mu$. Recall that, even if the initial $\rho$
is normalized to unit trace (as we always assume), the trace of
$\rho_\mu$ in the above equations is not equal to~1. Rather, it is the
probability of occurrence of outcome $\mu$. It is quite convenient to
keep $\rho_\mu$ unnormalized, with the above interpretation for its
trace.

The results obtained here are more general than those of Kraus, because
the matrices $A_{\mu m}$ may be rectangular. As \Eq{AU}) shows, these
matrices are simply related to the unitary transformation $U_{s\mu\sigma
m}$ that generates the premeasurement~\cite{ben}. Super\-operators that
do not conserve the number of dimensions of the density matrix were also
considered by other authors~\cite{Ben}. The present treatment is even
more general, because it allows the number of rows in $A_{\mu m}$ (that
is, the order of $\rho'_\mu$) to depend on $\mu$, since we may decide to
discard different subsystems according to the outcome of the
measurement. 

{}From \Eq{AU}), which relates the matrix elements $(A_{\mu m})_{\sigma
s}$ to the unitary transformation involved in the quantum intervention,
it appears that if we multiply the order of $\rho'_\mu$ by the range
of the indices $m$ in $A_{\mu m}$, the product of these two numbers is
the same for all $\mu$, since it is equal to the number of dimensions of
the composite system \cC, namely the original quantum system together
with the measuring apparatus. However, the situation is more
complicated, because it may happen that $A_{\mu m}=0$ for some values of
$\mu m$. Moreover, if the matrices $A_{\mu m}$ and $A_{\mu n}$ are
proportional to each other for some $m$ and $n$, these matrices can be
combined into a single one; and conversely, any $A_{\mu m}$ can be split
into several ones which are proportional to each other. Therefore there
is no simple rule saying how many terms appear in the the sum in
\Eq{ArhoA}).

The probability of occurrence of outcome $\mu$ in a measurement is given
by \Eq{prob}) and it can now be written as

\beq  p_\mu=\sum_m\Tr(A_{\mu m}\,\rho\,A_{\mu m}^\dagger)=
  \Tr(E_\mu\,\rho).\eeq
The positive (that is, non-negative) operators 

\beq E_\mu=\sum_m A_{\mu m}^\dagger\,A_{\mu m}, \label{Emu}\eeq
whose dimensions are the same as those of the initial $\rho$, are
elements of a POVM and satisfy $\sum_\mu E_\mu={\bf1}$. Note that null
outcomes (i.e., no detection) have to be included in that sum. They
indeed are the most probable result in typical experimental setups. Yet,
even if no detector is excited, the intervention may affect the quantum
system~\cite{Dicke}, and the corresponding $A_{\mu m}$ are not trivial.
There may even be several distinct $A_{\mu m}$ for ``no detection,''
depending on the cause of the failure.

Conversely, given $E_\mu$ (a non-negative square matrix of order $k$) it
is always possible to split it in infinitely many ways as in
\Eq{Emu}). This is easily proved by taking a basis in which $E_\mu$ is
diagonal. All the elements are non-negative, so that by taking their
square roots we obtain a matrix $\sqrt{E_\mu}$ that satisfies the
relation required for $A_{\mu m}$. Next, let $\{S_{\mu m}\}$ be a set
of complex rectangular matrices with $k$ columns and any number of rows,
satisfying $\sum_m S_{\mu m}^\dagger S_{\mu m}={\bf1}$. It follows that
$A_{\mu m}=S_{\mu m}\sqrt{E_\mu}$ satisfies \Eq{Emu}).

Moreover, if a POVM is factorable, namely

\beq E_{\mu\nu}=E_\mu^{(1)}\0E_\nu^{(2)}, \label{Efact} \eeq
where the indices (1) and (2) refer to two distinct subsystems, then
the above construction provides factorable Kraus matrices, 

\beq A_{\mu\nu mn}=A_{\mu m}^{(1)}\0A_{\nu n}^{(2)}. \eeq
The operator sum in Eqs.~(\ref{ArhoA}) and (\ref{Emu}) now becomes
double sums, over the indices $m$ and $n$. Such double sums are indeed
needed. If we had simply written, instead of \Eq{Efact}),
$E_{\mu}=E_\mu^{(1)}\0E_\mu^{(2)}$, the corresponding Kraus matrices
would in general not be factorable. Such a POVM, with a single index, is
called {\it separable\/} and it cannot in general be implemented by
separate operations on the two subsystems with classical communication
between them~\cite{Ben}.

The factorization of a POVM as in \Eq{Efact}) is not the most general
one. It corresponds to two POVMs independently chosen by the two
observers. However, the observers may also follow an adaptive stategy.
After the first one (conventionally called Alice) executes the POVM
$\{E_\mu^{(1)}$\}, she informs the second observer (Bob) of the result,
$\mu$ say, and then Bob uses a POVM adapted to that result. It will be
denoted as $\{E_{\nu\mu}^{(2)}$\}, with

\beq \sum_\nu E_{\nu\mu}^{(2)}={\bf1}^{(2)}\qquad\forall\mu. \eeq
Note that the chronological order of the Greek indices indicating the
outcomes of consecutive measurements is from right to left, just as the
order for consecutive applications of a product of linear operators. We
then have, instead of \Eq{Efact}),

\beq E_{\nu\mu}=E_\mu^{(1)}\0E_{\nu\mu}^{(2)}, \label{EEE}\eeq
whence

\beq A_{\nu\mu nm}=A_{\mu m}^{(1)}\0A_{\nu\mu n}^{(2)}. \label{AAA}\eeq

A more complicated situation arises when the {\it same\/} system or
subsytem is subjected to consecutive interventions that depend on the
outcomes of preceding interventions. We have, subsequent to the map in
\Eq{ArhoA}),

\beq \rho'_\mu\to\rho''_{\nu\mu}=\sum_n B_{\nu\mu n}\,\rho'_\mu\,
 B_{\nu\mu n}^\dagger.\eeq
The result of these two consecutive maps can be written

\beq \rho\to\rho''_{\nu\mu}=\sum_{n,m} C_{\nu\mu nm}\,\rho\,
 C_{\nu\mu nm}^\dagger,\eeq
where

\beq C_{\nu\mu nm}=B_{\nu\mu n}\,A_{\mu m}.\eeq
It follows that

\beq E_{\nu\mu}=\sum_{n,m} C_{\nu\mu nm}^\dagger\,C_{\nu\mu nm}=
 \sum_m A_{\mu m}^\dagger \Bigl(\sum_n B_{\nu\mu n}^\dagger\,
 B_{\nu\mu n}\,\Bigr)\,A_{\mu m}. \eeq

There is no simple relationship between this expression and the
antecedent POVM element $E_\mu$, unless the Kraus matrices $B_{\nu\mu
n}$ are chosen in such a way that

\beq \sum_{\nu,n}B_{\nu\mu n}^\dagger\,B_{\nu\mu n}={\bf1}.\eeq
We then have

\beq \sum_\nu E_{\nu\mu}=E_\mu. \eeq
This splitting of $E_\mu$ into a sum of several parts is called a POVM
{\it refinement\/}~\cite{PW91}. It may be repeated many times, until all
the final POVM elements are matrices of rank~1. For example, this can be
done by an apparatus that includes a multidimensional ancilla which is
{\it not\/} discarded at intermediate stages but only after the
completion of the measuring procedure. This is why it is convenient to
consider that ancilla explicitly rather than as part of the measuring
apparatus. In the resulting Neumark extension (that is, the joint
Hilbert space of the quantum system and the ancilla) each POVM is
implemented as an ordinary von Neumann PVM \cite{Hel,Neu}. Initially,
the latter is coarse grained: it distinguishes only multidimensional
subspaces of the system and the ancilla. This PVM is then gradually
``refined'' by the observer who uses further PVMs to select smaller and
smaller orthogonal subspaces. How these subspaces are explicitly defined
depends on results obtained in preceding tests. This is an adaptive
strategy which is particularly efficient for the optimal identification
of an unknown bipartite quantum state~\cite{PW91}. The two observers
take turns in performing local measurements and informing each other of
the results they obtained. The final result has again the form
(\ref{EEE}), but now the label $\mu$ stands for the entire sequence of
intermediate outcomes that were obtained by the two observers, and the
label $\nu$ indicates the result of the very last intervention.

For example, if each observer performs just two tests, with consecutive
results $\mu$ (Alice), $\nu$ (Bob), $\sigma$ (Alice), and $\tau$ (Bob),
then \Eq{AAA}) becomes

\beq A_{\tau\sigma\nu\mu tsnm}=
 A^{(1)}_{\nu\mu\sigma s}\,A^{(1)}_{\mu m}\0
 A^{(2)}_{\tau\sigma\nu\mu t}\,A^{(2)}_{\nu\mu n}.\eeq
This relationship is valid for any two pairs of consecutive tests, not
only for those of the ``refinement'' type.

Returning to the case of a single observer, let a complete set of POVM
elements be given. It is then always possible to construct from their
Kraus matrices (in infinitely many ways) unitary transformations that
satisfy \Eq{unitU}). Indeed, writing all indices explicitly, the
relation $\sum_\mu E_\mu={\bf1}$ becomes

\beq \sum_{\mu,m,\sigma}
  (A_{\mu m})_{\sigma s}^*\,(A_{\mu m})_{\sigma t}=\delta_{st}.\eeq
This can be written, owing to \Eq{AU}), as

\beq \sum_{\mu,m,\sigma}
  U_{s\mu\sigma m}^*\,U_{t\mu\sigma m}=\delta_{st},\eeq
which is the same as \Eq{unitU}) with the index $\lambda$ (which refers
to the composite system~\cC) replaced by the composite index
$s\mu\sigma$. This explicitly shows how any POVM can be implemented by a
unitary transformation and a suitable apparatus having the necessary
dimensionality (usually much smaller than the one required by the
introduction of an ancilla).

Recall that the $A_{\mu m}$ matrices are in general rectangular. For
example in the teleportation scenario~\cite{telep} Alice can discard her
two quantum particles together with her measuring apparatus after she
performs her 4-way test. The Hilbert space of Alice's subsystem then
becomes trivial: it has only one dimension (only one state). In that
case, $A_{\mu m}$ has a single row, $A_{\mu m}^\dagger$ a single column,
and $\rho_\mu$ is just a number, namely the probability of obtaining
outcome $\mu$. Likewise, each stage of a quantum
distillation~\cite{distil1,distil2} causes a reduction of the number of
dimensions of the quantum system that is distilled. That system
initially consists of a set of entangled subsystems. Successive
interventions select suitable subsets that have higher degrees of
entanglement. Ideally, the final result should be arbitrarily close to a
pair of spin-$1\over2$ particles in a singlet state.

Some authors consider only square matrices $A_{\mu m}$ and then it is
mathematically permitted to sum all the $\rho_\mu$ so as to obtain the
average state of all the outgoing quantum systems. For example, if the
outcomes of our interventions are not recorded, so that no subensembles
are selected, we may write

\beq  \rho'=\sum_{\mu,m} A_{\mu m}\,\rho\,A_{\mu m}^\dagger,
  \label{noselect} \eeq
where the trace of $\rho'$ is 1, of course. Such sums are rarely needed
in theoretical discussions. Different labels $\mu$ correspond to
different world histories (that is, different samples in the ensemble of
experiments). Summing over them is like saying that peas and peanuts
contain on the average 42\% of water, instead of saying that peas have
78\% and peanuts 6\%~\cite{USDA}. Still, there are some cases where this
kind of averaging is justified. For example, when we compare the
expected yields of various distillation methods~\cite{distil1,distil2},
we are interested only in average results. Moreover, when the quantum
system weakly interacts with an unknown environment (such as a heat
bath), rather than with an apparatus that can neatly distinguish
different outcomes of the intervention, the result is a continuous
decoherence of the quantum state~\cite{Zurek}. This issue is discussed
quantitatively in the next section.

\bc {\bf V. THE LINDBLAD EQUATION}\ec

For a complete treatment of the quantum system in its unknown
environment, we write the Hamiltonian as

\beq H=H_0+H_{{\rm env}}+H_{{\rm int}}, \eeq
with obvious notations. The last two terms in $H$ generate a stochastic,
rapidly fluctuating motion. The exact evolution, taking everything into
account, is a Brownian motion (a kind of random walk) superimposed on
the ideal motion. Assume that the unperturbed evolution due to $H_0$ is
very slow on the scale of $t_{{\rm decoh}}$, the time needed for
\Eq{decohe}) to be valid. One can thus write (combining the composite
index $\mu m$ into a single index $j$)

\beq A_j = S_j + F_j, \eeq
where the matrices $S_j$ correspond to the slow motion generated by
$H_0$, and the matrices $F_j$ to the fast fluctuations due to the
environment. This neat decomposition into slow and fast variables
involves of course some arbitrariness that will be reflected in the
derivation of the Lindblad equation below.

It follows from Eq.~(\theequation) that $A_j\rho A_j^\dagger$ splits
into three kinds of terms. Those quadratic in $S_j$ represent the smooth
evolution due to $H_0$. If we wish to write a differential equation for
$d\rho/dt$, the other terms have to be smoothed out on a time scale much
longer than $t_{{\rm decoh}}$. Since this is a random walk, the terms
linear in $F_j$ average out to zero on that time scale, and the terms
that are quadratic in $F_j$ grow linearly in time. Clearly, this
smoothing out and the resulting linear growth involve some
approximations whose validity has to be ascertained on a case by case
basis.

Let us thus assume that there is a coarse time scale $\delta t\gg
t_{{\rm decoh}}$, long enough so that the fluctuations are averaged out,
and yet short enough so that the slow evolution due to $H_0$ is
negligible beyond first order in $\delta t$. We can then write
$F_j\simeq V_j\sqrt{\delta t}$, where the matrices $V_j$ are finite. The
result is the Lindblad equation~\cite{lindblad}

\beq d\rho/dt=i[\rho,H_0]+\sum_j(V_j\rho V_j^\dagger
 -\mbox{$1\over2$}\rho V_j^\dagger V_j
 -\mbox{$1\over2$}V_j^\dagger V_j\rho). \label{QDS} \eeq
This equation is of course valid only in the future time direction
($dt>0$), because the smoothing out of fluctuations entails an
irreversible loss of information. Lindblad's original derivation used an
abstract argument involving complete positivity and a semigroup
structure (again $dt>0$). An equivalent argument was given independently
by Gorini {\it et al.\/}~\cite{gorini}. The present proof is based on an
explicit dynamical model of interaction and it may be easier to
understand. A similar derivation was also obtained by
Schumacher~\cite{schu} and, after the present article was submitted for
publication, a more detailed discussion, also based on the Kraus
formalism, was published by Bacon {\it et al.\/}~\cite{bacon}. Still
more recently, Adler sharpened the above heuristic derivation of
\Eq{QDS}) by using It\^o's stochastic calculus~\cite{adler}.\bigskip

\bc {\bf ACKNOWLEDGMENTS}\ec

I am grateful to Chris Fuchs, Ady Mann, Ben Schumacher, Barbara Terhal,
and Daniel Terno for many helpful comments. This work was supported by
the Gerard Swope Fund and the Fund for Encouragement of Research.

\end{document}